\documentclass[letterpaper,10pt,twocolumn]{article}
\usepackage[T1]{fontenc}
\usepackage{parskip}
\usepackage[font={small}]{caption}
\usepackage[hidelinks]{hyperref}
\usepackage[margin=2cm]{geometry}
\usepackage{times}
\usepackage{amsmath,amssymb,amsthm,amsfonts}
\usepackage[affil-it]{authblk}
\usepackage{listings}

\usepackage{graphicx}      
\usepackage[round]{natbib}        
\usepackage{subcaption}
\usepackage{dblfloatfix}
\usepackage{algorithm}
\usepackage{algorithmic}

\usepackage{sectsty}
\sectionfont{\large}
\subsectionfont{\normalsize}

\usepackage[compact]{titlesec}
\titlespacing{\section}{0pt}{*0}{*0}
\titlespacing{\subsection}{0pt}{*0}{*0}

\makeatletter
\newcommand{\settitle}{\@maketitle}
\makeatother

\usepackage[outline]{contour}% http://ctan.org/pkg/contour
\usepackage{xcolor}

\newtheorem*{remark}{Remark}

\date{}
 
\title{Improving the particle filter in high dimensions using conjugate artificial process noise}
\author[]{Anna Wigren} 
\author[]{Lawrence Murray} 
\author[]{Fredrik Lindsten}

\affil[]{\normalsize Department of Information Technology, Uppsala University, Sweden \\ \normalsize Email: \{anna.wigren, lawrence.murray, fredrik.lindsten\}@it.uu.se}

\begin{document}

\onecolumn 

\settitle
\thispagestyle{empty}
\addtocounter{page}{-1}

\textbf{Please cite this version:}

Anna Wigren, Lawrence Murray, Fredrik Lindsten. Improving the particle filter in high dimensions using conjugate artificial process noise. In \textit{Proceedings of the 18th IFAC Symposium on System Identification (SYSID)},
Stockholm, Sweden, 2018.

\begin{center}
\begin{minipage}{.75\linewidth}
\begin{lstlisting}[breaklines,basicstyle=\small\ttfamily]
	@inproceedings{AWigren2018,
	title = "Improving the particle filter in high dimensions using conjugate artificial process noise",
	author = "Anna Wigren and Lawrence Murray and Fredrik Lindsten",
	booktitle = "Proceedings of the 18th IFAC Symposium on System Identification (SYSID)",
	pages = "670 - 675",
	year = "2018",
	address = "Stockholm, Sweden",
	}	
\end{lstlisting}
\end{minipage}
\end{center}

\vspace{5em}

\newpage	

\twocolumn	
	
\twocolumn[
\begin{@twocolumnfalse}	

	\maketitle
	\vspace{-.3cm}
	\begin{abstract}                
		The particle filter is one of the most successful methods for state inference and identification of general non-linear and non-Gaussian models. However, standard particle filters suffer from degeneracy of the particle weights, in particular for high-dimensional problems. We propose a method for improving the performance of the particle filter for certain challenging state space models, with implications for high-dimensional inference. First we approximate the model by adding artificial process noise in an additional state update, then we design a proposal that combines the standard and the locally optimal proposal. This results in a bias-variance trade-off, where adding more noise reduces the variance of the estimate but increases the model bias. The performance of the proposed method is empirically evaluated on a linear-Gaussian state space model and on the non-linear Lorenz'96 model. For both models we observe a significant improvement in performance over the standard particle filter. 
	\end{abstract}
	\vspace{.5cm}

\end{@twocolumnfalse}
]	

\section{Introduction}
Non-linear and high-dimensional state space models arise in many areas of application, such as oceanography \citep{Mattern13}, numerical weather prediction \citep{Evensen94} and epidemiology \citep{ He10,Shaman12}, to mention a few. Here we consider models of the form
\begin{equation} \label{eq:Omod}
\begin{aligned}
x_{t}&=f(x_{t-1},v_{t}) \\
y_{t}&=Cx_{t}+e_{t}						
\end{aligned}
\end{equation}
where $f$ is a non-linear function of the previous state $x_{t-1}$, $v_{t}$ is process noise, and the observations $y_{t}$ are a linear function of the current state $x_{t}$ with additive Gaussian noise $e_{t}\sim \mathcal{N}(0,R)$. This form of the non-linear state dynamics $f(x_{t-1},v_{t})$ is very general compared, e.g., to the case when the process noise is just additive, and allows for using blackbox simulation models, discretized stochastic differential equations, etc. For the observations we restrict our attention to the linear-Gaussian case, which is common in many applications. However, the method we propose can handle any observation likelihood for which there is a conjugate prior. One such case is a Poisson distributed observation likelihood (with a Gamma distributed prior). 

When performing filtering we wish to recover the unknown states $x_{t}$ at time $t$ given all observations $y_{1:t}$ of the process up to time $t$. The filtering distribution $p(x_{t}|y_{1:t})$ can only be evaluated in closed form for a few specific models, such as the linear-Gaussian state space model (the Kalman filter). In more general cases the filtering distribution must be approximated. The particle filter is one way to do this by representing the filtering distribution with a set of weighted samples from the distribution. 

In addition to being of significant interest on its own, the filtering problem is also intimately related to model identification via both maximum likelihood and Bayesian formulations (see, e.g., \citet{Schon15}). Indeed, the data log-likelihood can be expressed as a sum of filtering expectations. Thus, even though we will restrict our attention to the filtering problem, we emphasize that the improvements offered by the new method are useful also for identification of models of the form \eqref{eq:Omod}.

The particle filter can, unlike the Kalman filter, handle highly non-linear models, but may also experience degeneracy of the particle weights. Degeneracy occurs when one particle has a weight close to one while the weights of all other particles are close to zero. The filtering distribution is then effectively represented by a single particle, which results in a very poor approximation. It has been shown that to avoid weight degeneracy the number of particles must increase exponentially with the state dimension \citep{Snyder15}. Weight degeneracy is therefore a frequent issue for high-dimensional problems. 

A range of different techniques for high-dimensional filtering have been previously developed. Methods like the ensemble Kalman filter \citep{Evensen94} can be used to solve the filtering problem if the system is mildly non-linear. For more difficult cases adaptation of the particle filter to higher dimensions is necessary. Some examples of particle filters for high-dimensional problems include \citet{Djuric13,Naesseth15,Rebeschini15, Kunsch17}. These methods all aim to approximate the particle filter algorithm in some sense to avoid degeneracy. The method we propose here is in a different vein---we will instead approximate the model \eqref{eq:Omod} by adding artificial process noise. The filtering problem is then solved using a regular particle filter with the proposal chosen as a combination of the standard (bootstrap) proposal and the locally optimal proposal. This is related to "roughening", first introduced by \citet{Gordon93}, where artificial noise is added after resampling to spread the particles in an attempt to mitigate degeneracy. Here we refine this concept by proposing a specific proposal for the approximate model to improve the performance for high-dimensional models. Based on results by \citet{Snyder15}, we also provide insights on how approximating the model by adding noise can be seen as a bias-variance trade-off where the magnitude of the artificial process noise is a tuning parameter.

\section{Background on the particle filter} 

The particle filter sequentially approximates the filtering distribution  as $\hat{p}^{N}(x_{t}|y_{1:t})=\sum_{i=1}^{N}w_{t}^{i}\delta_{x_{t}^{i}}(x_{t})$ where $x_{t}^{i}$ are random samples (particles), $w_{t}^{i}$ are their corresponding weights, $\delta$ is the Dirac delta function and $N$ is the number of particles. It is often impossible to sample from the filtering distribution directly, instead importance sampling is used where samples are drawn sequentially from a proposal distribution $q(x_{t}|x_{t-1},y_{t})$. The proposal can be any distribution from which it is possible to draw samples and for which $q(x_{t})>0$ whenever $p(x_{t})>0$.         
To adjust for not sampling from $p$ a correction is introduced in the weight update. The unnormalized importance weights are given by
\begin{equation} \label{eq:wProp}
\begin{aligned}
\tilde{w}_{t}^{i}&\propto \frac{p(x_{t}^{i},x_{t-1}^{i}|y_{1:t})}{q(x_{t}^{i},x_{t-1}^{i}|y_{1:t})} \propto \frac{p(x_{t}^{i}|x_{t-1}^{i})p(y_{t}|x_{t}^{i})}{q(x_{t}^{i}|x_{t-1}^{i},y_{t})}w_{t-1}^{i}
\end{aligned}
\end{equation}
where $w_{t-1}^{i}$ is the normalized weight from the previous time step. The normalized importance weights are $w_{t}^{i}=\tilde{w}_{t}^{i} / \sum_{i=1}^{N}\tilde{w}_{t}^{i}$.

Each iteration in the filtering algorithm consists of three steps. First resampling is (possibly) performed according to the normalized weights $w_{t-1}$ from the previous time step, and the weights are set to $1/N$. The particles are then propagated to the next time step using the proposal distribution $q$. Finally the normalized importance weights are computed as described above. Further details on the particle filtering algorithm can be found e.g. in \citep{Doucet00}.

\subsection{The standard proposal}
A common choice of proposal distribution is the transition density $p(x_{t}|x_{t-1})$, referred to as the standard proposal. Inserting this proposal in \eqref{eq:wProp} gives the unnormalized weights $\tilde{w}_{t}=p(y_{t}|x_{t})w_{t-1}$. If resampling is performed in every iteration this choice of proposal corresponds to the original version of the particle filter, the bootstrap filter \citep{Gordon93}. Note that it is sufficient to be able to simulate from $p(x_{t}|x_{t-1})$, exact evaluation of the expression is not required. Therefore the standard proposal can be used even for models like \eqref{eq:Omod} where, typically, no closed form expression is available for the transition density. Furthermore, the corresponding weights, given by the observation likelihood, can be easily evaluated. However, this choice of proposal is prone to weight degeneracy, in particular when the system is high-dimensional or when the observations are very informative (low observation noise) or contain outliers \citep{Cappe07}. This degeneracy occurs when there is little overlap between the observation likelihood and the prior distribution of particles.

\subsection{The locally optimal proposal}
A possible remedy for the shortcomings of the standard proposal is to use a proposal which shifts the particles towards the observations by taking both the previous state $x_{t-1}$ and the current observation $y_{t}$ into account when propagating and reweighting the particles. One such choice is the locally optimal proposal, which is optimal in the sense that the variance of the importance weights is minimized when compared to other proposals depending only on $x_{t-1}$ and $y_{t}$ \citep{Doucet00}. The locally optimal proposal propagates the particles according to 
\begin{equation} \label{eq:Optq}
q(x_{t}|x_{t-1},y_{t})=p(x_{t}|x_{t-1},y_{t})=\frac{p(x_{t}|x_{t-1})p(y_{t}|x_{t})}{p(y_{t}|x_{t-1})}
\end{equation} 
and then reweights using the importance weights $\tilde{w}_{t}^{i}=p(y_{t}|x_{t-1}^{i})w_{t-1}^{i}$. Unfortunately it is often not possible to use this proposal due to two major difficulties; it must be possible both to sample from the proposal $p(x_{t}|x_{t-1},y_{t})$ and to evaluate $p(y_{t}|x_{t-1})=\int p(y_{t}|x_{t})p(x_{t}|x_{t-1})dx_{t}$. This integral can, in most cases, only be evaluated when $p(y_{t}|x_{t})$ is conjugate to $p(x_{t}|x_{t-1})$.    

The locally optimal proposal is in general not available for the model \eqref{eq:Omod}. One exception is the special case when the state dynamics are non-linear with additive Gaussian noise, that is when $x_{t}=f(x_{t-1})+v_{t}$ \citep{Doucet00}. Assuming $v_{t}$ and $e_{t}$ are independent Gaussian noise with mean zero and covariances $Q$ and $R$ respectively, \eqref{eq:Omod} can be expressed using the densities
\begin{equation}
%\begin{aligned}
x_{t}|x_{t-1}\sim \mathcal{N}(f(x_{t-1}),Q), \hspace{2mm} y_{t}|x_{t}\sim \mathcal{N}(Cx_{t},R).
%\end{aligned}
\end{equation}
Both densities are Gaussian, so well-known relations give the proposal $p(x_{t}|x_{t-1},y_{t})=\mathcal{N}(x_{t}|\mu_{c},\Sigma_{c})$ where
\begin{equation} \label{eq:optLGq}
\begin{aligned}
\mu_{c}&=f(x_{t-1})+QC^{T}(R+CQC^{T})^{-1}(y_{t}-Cf(x_{t-1})) \\ 
\Sigma_{c}&=Q-QC^{T}(R+CQC^{T})^{-1}CQ,
\end{aligned}
\end{equation}  
and for the corresponding weights we obtain
\begin{equation} \label{eq:optLGw}
p(y_{t}|x_{t-1})=\mathcal{N}(y_{t}|Cf(x_{t-1}),R+CQC^{T}).
\end{equation}

\section{Particle filter with conjugate artificial process noise}
The locally optimal proposal has minimal degeneracy compared to other proposals, and for high-dimensional systems it can improve upon the standard proposal by several orders of magnitude \citep{Snyder15}. Unfortunately, as pointed out in the previous section the locally optimal proposal is in general not available for \eqref{eq:Omod} and common approximations, e.g. based on local linearizations \citep{Doucet00}, are not applicable when the transition density function is intractable. However, to still be able to leverage the benefits of the locally optimal proposal we propose a controlled approximation of \eqref{eq:Omod} where artificial noise is added in an extra state update. The approximate model is given by
\begin{subequations} \label{eq:ANmod}
	\begin{align}
	x_{t}'&= f(x_{t-1},v_{t}) \label{eq:ANmod_a} \\ 
	x_{t} &= x_{t}'+\varepsilon \xi_{t}  \label{eq:ANmod_b}  \\
	y_{t} &= Cx_{t}+e_{t}  \label{eq:ANmod_c}
	\end{align}
\end{subequations}
where $\varepsilon$ is a parameter adjusting the magnitude of the artificial noise. We consider a linear-Gaussian observation model \eqref{eq:ANmod_c}, hence for conjugacy between \eqref{eq:ANmod_b} and \eqref{eq:ANmod_c} $\xi_{t} \sim \mathcal{N}(0,S)$ where $S$ is a covariance matrix. Note that by choosing $\varepsilon=0$ we recover the original model \eqref{eq:Omod}.

To design a particle filter for \eqref{eq:ANmod} we must choose a proposal and derive the corresponding weights. If $x_{t}'$ and $x_{t}$ are taken to be the system states, the model \eqref{eq:ANmod} suggests using a combination of the standard and the locally optimal proposal. First the particles are propagated according to the standard proposal \eqref{eq:ANmod_a}. Noting that the two latter equations \eqref{eq:ANmod_b} and \eqref{eq:ANmod_c} are linear-Gaussian the particles are then propagated according to the locally optimal proposal taking $x_{t}'$ to be the previous state. Using  \eqref{eq:optLGq} and \eqref{eq:optLGw} we obtain the combined proposal 
\begin{equation} \label{eq:propAN}
\begin{aligned}
&q(x_{t},x_{t}'|x_{t-1},y_{t})=p(x_{t}|x_{t}',y_{t})p(x_{t}'|x_{t-1}) \\
&\hspace{2.5cm}=\mathcal{N}(x_{t}|\mu, \Sigma)p(x_{t}'|x_{t-1}) \\
&\hspace{0.05cm}\mu=x_{t}'+\varepsilon^{2}S C^{T}(R+C\varepsilon^2S C^{T})^{-1}(y_{t}-Cx_{t}') \\
&\Sigma=\varepsilon^{2}S-\varepsilon^{2}SC^{T}(R+C\varepsilon^2SC^{T})^{-1}C\varepsilon^{2}S.
\end{aligned}
\end{equation} 
where $p(x_{t}'|x_{t-1})$ is the standard proposal \eqref{eq:ANmod_a}. The corresponding importance weights are
\begin{equation} \label{eq:wAN}
\tilde{w}_{t}=p(y_{t}|x_{t}')w_{t-1}=\mathcal{N}(y_{t}|Cx_{t}',R+C\varepsilon^{2}SC^{T})w_{t-1}.
\end{equation}
It is clear from \eqref{eq:propAN} and \eqref{eq:wAN} that choosing $\varepsilon=0$ will recover the standard proposal. Equation \eqref{eq:wAN} also indicates why this choice of proposal is beneficial for high-dimensional problems; adding artificial process noise ($\varepsilon >0$) will make the covariance parameter larger which in turn will make the weights less degenerate.

One way to interpret the proposed strategy is that adding artificial process noise in \eqref{eq:ANmod} will introduce some extra movement to the particles. The first propagation (standard proposal) moves the initial particles $x_{t-1}^{i}$ according to the state dynamics. If we add noise ($\varepsilon>0$) the optimal proposal then moves the propagated particles further, shifting them towards the observation $y_t$ according to the expression for the mean in \eqref{eq:propAN}.

\citet{Snyder15} found that the difference in performance between the locally optimal and the standard proposal increases with the magnitude of the process noise. Effectively this means that when using the locally optimal proposal the Monte Carlo variance of e.g. estimates of the normalizing constant or test functions, such as the posterior mean or variance of the system state, can be reduced by adding more process noise. However, adding more artificial process noise in \eqref{eq:ANmod} will introduce more bias in our estimates, so ultimately our proposed method has a bias-variance trade-off where $\varepsilon$ is the tuning parameter.

\subsection{Choice of parameters}
The parameter $\varepsilon$ adjusts the magnitude of the noise and hence controls the bias-variance trade-off. A high value on $\varepsilon$ implies that there is a lot of freedom in moving the particles in the second stage of the proposal which results in a lower Monte-Carlo variance, but at the cost of a higher model bias. 

For the case of a linear-Gaussian observation model the covariance matrix $S$, describing the correlation structure, must also be specified. A simple choice is to use the identity matrix which corresponds to adding noise of the same magnitude to all states, but with no correlation between states. However, if some states are not observed they will not be affected by the artificial process noise---they will just be propagated blindly forwards according to the standard proposal. Another possible choice is to use the weighted sample covariance matrix. This choice will take the correlation structure of the states into account which can mitigate the impact of not observing some states. Each element of the weighted sample covariance matrix $\Xi$ at time $t$ is given by
\begin{equation} \label{eq:SC}
\Xi_{jk}=\frac{1}{1-\sum_{i=1}^{N}(w^{i})^{2}}\sum_{i=1}^{N}w^{i}(x'^{i}_{j}-\mu_{j})(x'^{i}_{k}-\mu_{k})
\end{equation}  
where $j,k=1...d$, $d$ is the state dimension, $w^{i}$ is the normalized weight of particle $i$, $x'^{i}_{j}$ is the value of the $j\text{:th}$ dimension of particle $i$ and $\mu_{j}, \mu_{k}$ are the sample mean for dimension $j,k$ given by $\mu_{j}=\sum_{i=1}^{N}w^{i}x'^{i}_{j}$.

\section{Numerical examples}
To evaluate our proposed method we consider two examples; a linear-Gaussian state space model and the non-linear Lorenz'96 model \citep{Lorenz96}. For both models we examine the 10-dimensional case where only half of the states are observed (in noise) at each time step. Two choices of covariance matrices for the artificial process noise are considered; the block-diagonal matrix $B$ with an identity matrix in the upper block and zeros in the lower block, and the weighted sample covariance matrix $\Xi$ with elements given by \eqref{eq:SC}. The first choice will add artificial process noise to the observed states only, with no correlation between the states, whereas the latter choice will add artificial process noise to all states and allows for correlation between the states.

% LINEAR
\begin{figure*}[t] %
	\begin{center}
		\begin{subfigure}{\columnwidth}
			\includegraphics[width=8.4cm]{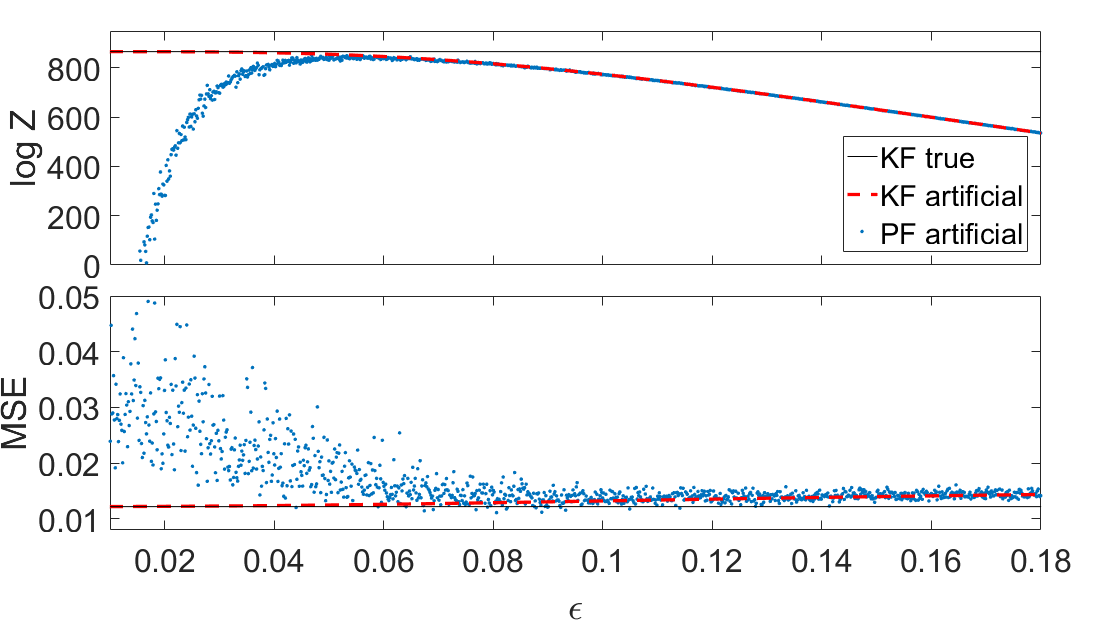}    
		\end{subfigure}\hfill%
		\begin{subfigure}{\columnwidth}
			\includegraphics[width=8.4cm]{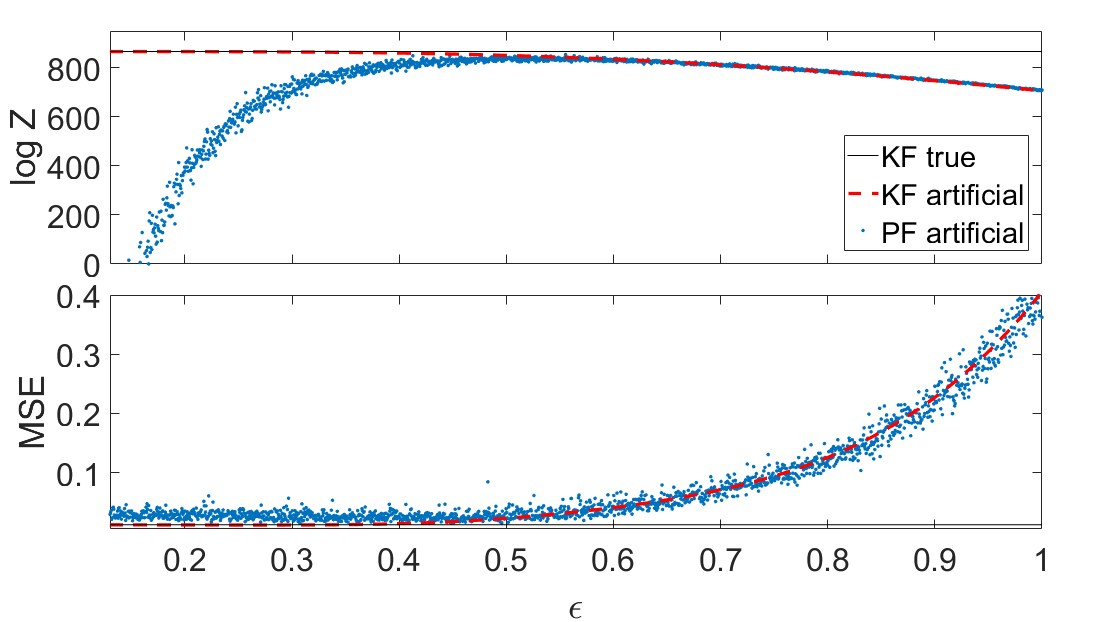}    
		\end{subfigure}\hfill%
		\caption{Marginal log-likelihood and MSE as a function of $\varepsilon$ for a 10-dimensional linear-Gaussian state space model. Left: Block-diagonal covariance matrix. Right: Weighted sample covariance matrix. The black solid line is the true Kalman filter estimate, the dashed red line is the Kalman filter estimate for \eqref{eq:ANmod} and the blue dots are the particle filter estimates for \eqref{eq:ANmod}.}
		\label{fig:LLHMSE_Lin}
	\end{center}
\end{figure*}

To quantify the performance of our method we use two measures; the marginal log-likelihood of the data, $\log Z =\log p(y_{1:T})$, and the mean square error (MSE) of the state estimates averaged over all time steps and all dimensions. The likelihood of the data is of particular interest for parameter estimation and model checking, and the MSE of the state estimate is of interest for filtering applications. In our discussion of the performance we will also refer to the effective sample size (ESS) for the particle filter given by $N_{\text{eff}}=1/ \sum_{i=1}^{N} (w_{t-1}^{i})^{2}$. If the ESS drops too low at some point the filter estimates will be degenerate since all particles will originate from a small number of ancestor particles.

\subsection{Linear Gaussian model} 
We first consider a 10-dimensional linear state space model with Gaussian noise of the form 
\begin{equation} \label{eq:LGSSM}
x_{t}=Ax_{t-1}+v_{t}, \hspace{5mm} y_{t}=Cx_{t}+e_{t}
\end{equation}
where $A$ is tridiagonal with value $0.6$ on the diagonal, $0.2$ on the first diagonal above and below giving a local dependence between the states, and zeros everywhere else. We observe half of the states, hence $C$ is $5\times 10$ where the left half is an identity matrix and the right half is a zero matrix. To make the estimation problem harder we assume that the covariance of the measurement noise is two orders of magnitude smaller than the covariance of the process noise ($10^{-4}$ vs $10^{-2}$).
Data for $T=200$ time steps was generated from \eqref{eq:LGSSM} and $N=1000$ particles were used to estimate the states with the particle filter given by \eqref{eq:propAN} and \eqref{eq:wAN}. 

For the linear-Gaussian state space model the optimal solution to the filtering problem is available from the Kalman filter, hence it is possible to compare the performance of our method with the best achievable performance. We will compare both with the true Kalman filter estimate and with the Kalman filter estimate for the approximate model~\eqref{eq:ANmod}. 

Fig. \ref{fig:LLHMSE_Lin} shows the log-likelihood and the MSE as a function of $\varepsilon$ for $S=B$ (left) and $S=\Xi$ (right). It is clear from the log-likelihood plots that the standard proposal ($\varepsilon=0$) degenerates whereas with our proposed method it is possible to almost reach the log-likelihood and MSE for the true Kalman filter for certain ranges of $\varepsilon$. 
It is also evident that there is a bias-variance trade-off with a higher variance for low values on $\varepsilon$ and a lower variance but bigger bias for larger values on $\varepsilon$.

\begin{remark} 
	Note that for small values of $\varepsilon$ the negative bias in the estimate of $\log Z$ is an effect of the increased Monte Carlo variance. It is well known that the particle filter estimate of $Z$ is unbiased, which by Jensen's inequality implies $\mathbb{E}(\log \hat{Z}) \leq \log Z$, where a large Monte Carlo variance tends to result in a large negative bias. By a log-normal central limit theorem \citep{Berard2014} it holds that the bias is roughly $-\sigma^{2}/2$ for $N$ large enough, where $\sigma^{2}$ is the Monte Carlo variance.
\end{remark}

For the sample covariance the highest log-likelihood and lowest MSE is obtained for similar values on $\varepsilon$, around $0.5$. For the identity matrix on the other hand the range of values for $\varepsilon$ giving the highest log-likelihood are lower than the range of $\varepsilon$ giving the lowest MSE. As $\varepsilon$ increases both choices of covariance matrices approaches the Kalman filter estimate for the approximate model \eqref{eq:ANmod} corresponding to the best we can do.

% LORENZ ZIN
\begin{figure*}[t]%
	\centering
	\begin{subfigure}{\columnwidth}
		\includegraphics[width=8.4cm]{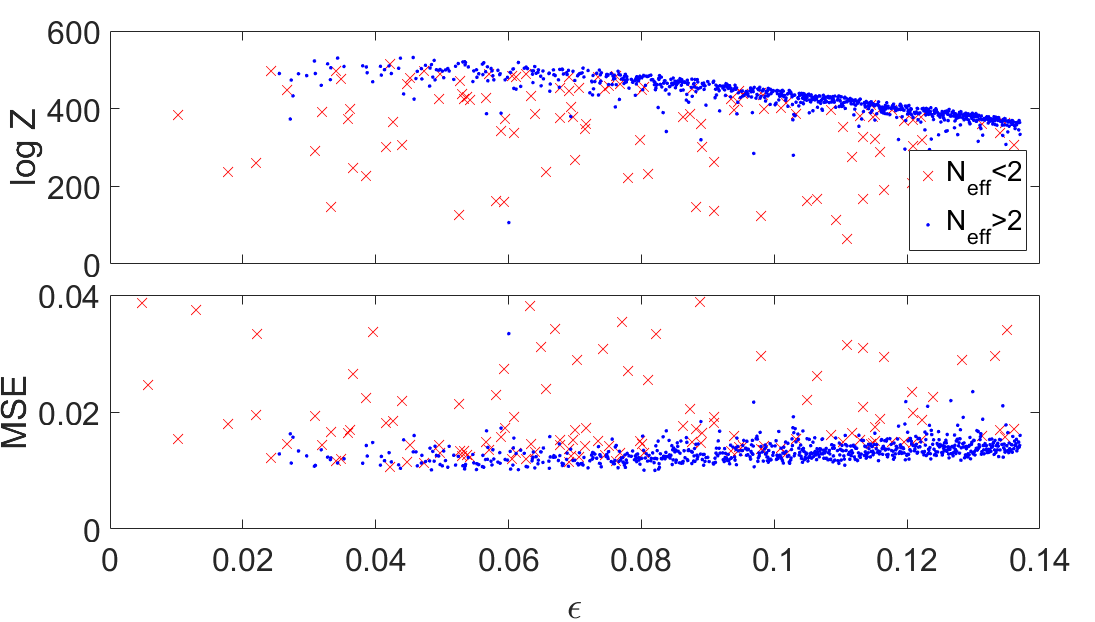}    	
	\end{subfigure}\hfill%
	\begin{subfigure}{\columnwidth}
		\includegraphics[width=8.4cm]{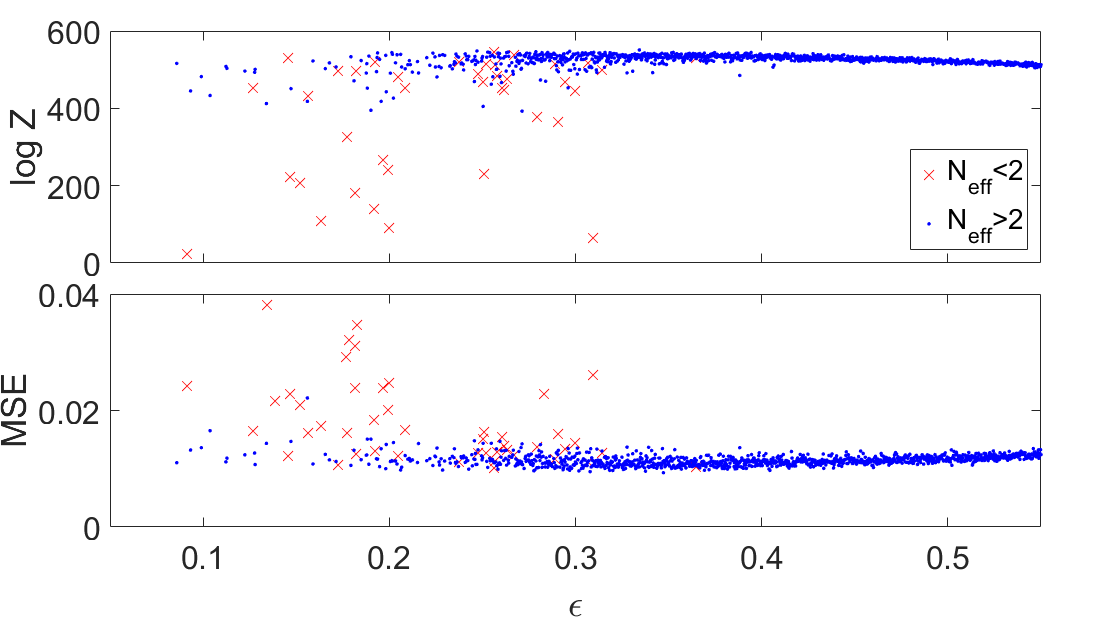}    
	\end{subfigure}\hfill%
	\caption{Marginal log-likelihood and MSE as a function of $\varepsilon$ for the 10-dimensional Lorenz'96 model, zoomed in on the choices of $\varepsilon$ which (mostly) avoids degeneracy. Left: Block-diagonal covariance matrix. Right: Weighted sample covariance matrix. Red crosses show estimates where the ESS drops below two at least once (indicating a degenerate filter estimate) and blue dots show estimates where the ESS is always greater than two.}
	\label{fig:LLHMSE_Lor_zin}
\end{figure*}

\begin{figure}[]%
	\centering
	\begin{subfigure}{.5\columnwidth}
		\includegraphics[width=4.2cm]{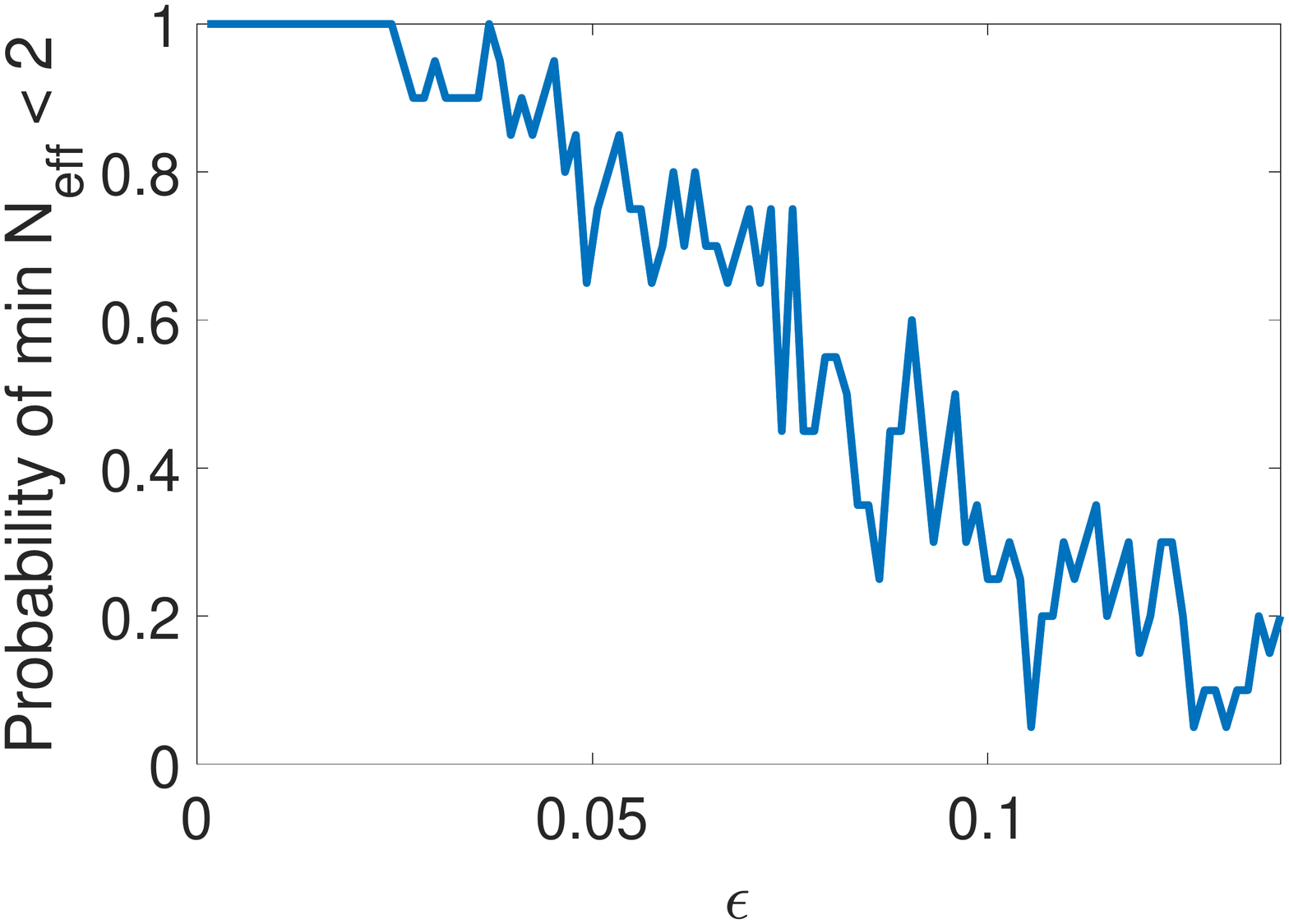}   
	\end{subfigure}\hfill%
	\begin{subfigure}{.5\columnwidth}
		\includegraphics[width=4.2cm]{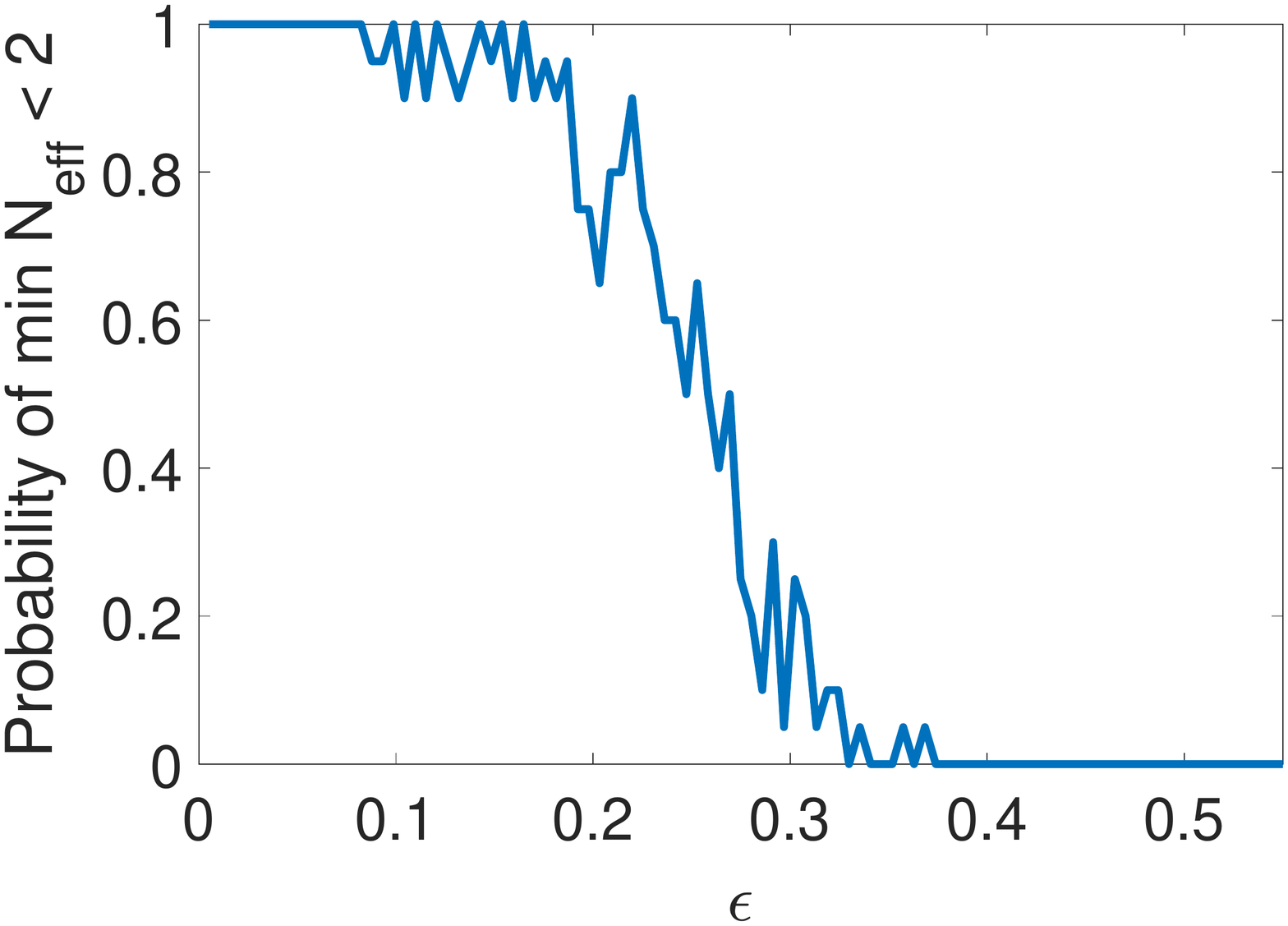}    
	\end{subfigure}\hfill%
	\caption{Estimated probabilities of degeneracy in terms of the ESS for varying $\varepsilon$. Left: Block-diagonal covariance matrix. Right: Weighted sample covariance matrix. }
	\label{fig:Neff}
\end{figure}

\subsection{Lorenz'96 model}
Next we consider the non-linear, possibly chaotic Lorenz'96 model which is often used for testing data assimilation algorithms \citep{Lorenz96}. The d-dimensional Lorenz'96 model is defined in terms of coupled stochastic differential equations which, for the continuous state $X(t)=( X_{1}(t), \dots, X_{d}(t) )^{T}$, are given by
\begin{align} \label{eq:Lor96}
\mathrm{d}X_{k}(t)=& \Big(\big(X_{k+1}(t)-X_{k-2}(t)\big)X_{k-1}(t)-X_{k}(t)+F\Big)\mathrm{d}t \nonumber \\ &+ b\mathrm{d}W_{k}(t),
\end{align}  
where the first term is drift and the second term is diffusion. The model is cyclic, hence $X_{-1}(t)=X_{d-1}(t)$, $X_{0}(t)=X_{d}(t)$ and $X_{d+1}(t)=X_{1}(t)$ is assumed. $F$ is a forcing constant confining the volume in which the solution can move and, for a fixed dimension $d$ of the state space, it determines whether the system exhibits chaotic, decaying or periodic behavior \citep{Karimi10}. We consider the 10-dimensional case and to obtain a highly non-linear model exhibiting chaotic behavior we use $F=12$. For the diffusion term we choose $b=0.1$ and $W(t)$ is a standard 10-dimensional Wiener process. The observations are linear-Gaussian and, like in \eqref{eq:LGSSM}, we observe only half of the states.  

Data was generated for $T=200$ steps assuming observations are made with a timestep $\Delta t=0.1$. The system can be discretized by considering the discrete state $x_{t}=X(t \Delta t)$ which, between observations, is propagated forward according to \eqref{eq:Lor96} using $M=15$ iterations of the Euler-Maruyama method for numerical solution of stochastic differential equations. Note that this system, unlike the previously considered linear-Gaussian system, is one example of \eqref{eq:Omod} where no closed form expression for $p(x_{t}|x_{t-1})$ exists which makes the filtering problem particularly challenging. For the artificial process noise particle filter we use $N=2000$ particles and the propagation of the states is a simulation forward in time using the Euler-Maruyama method for solving \eqref{eq:Lor96} numerically.

For small values of $\varepsilon$ the particle filter tends to degenerate, resulting in very poor estimates of the log-likelihood (as low as $-3\cdot 10^{6}$) and high MSE values (as high as 60). For clarity of presentation we therefore split the particle filter runs into two groups: one containing the degenerate cases, in which the ESS dropped below two at least once during the run, and one for the non-degenerate cases, in which the ESS stayed above two. Fig. \ref{fig:LLHMSE_Lor_zin} has been zoomed in on the non-degenerate runs, for $S=B$ (left) and $S=\Xi$ (right). Plots showing all runs are given in Fig. \ref{fig:LLHMSE_Lor_zo} in the appendix.

From Fig. \ref{fig:LLHMSE_Lor_zin} we observe that indeed there is a bias-variance trade-off for the proposed method where a higher value of $\varepsilon$ reduces the variance but increases the bias. A comparison of the log-likelihood plots for $S=B$ and $S=\Xi$ shows that the latter choice obtains higher likelihood estimates which decay more slowly and with a lower spread on the estimates for high values on $\varepsilon$. Similarly, the MSE estimates also show less spread for the choice $S=\Xi$. This indicates that the sample covariance matrix might give an estimate which is more robust for varying values of $\varepsilon$ for this model.

To examine the effect of $\varepsilon$ on the degeneracy of the particle filter in more detail we show the estimated probabilities of degeneracy (as defined above in terms of the ESS) for varying $\varepsilon$ in Fig. \ref{fig:Neff}, for $S=B$ (left) and $S=\Xi$ (right). These probabilities are estimated based on binning the range of $\varepsilon$ into 100 equally sized bins and counting the number of degenerate runs in each bin. This explains the raggedness of the estimates, but the plots nevertheless give an idea of how the probability of degeneracy decreases with increasing $\varepsilon$ (we expect that the true probabilities are monotonically decreasing).

\section{Conclusion}

The particle filter is a powerful inference method with strong convergence guarantees. However, for challenging cases such as high-dimensional models the well-known degeneracy issue of the particle filter can cause the Monte Carlo error to be significant, effectively rendering the standard particle filter useless. Furthermore, for many models of interest---in particular non-linear models of the form \eqref{eq:Omod}---it is difficult to improve over the standard particle filter proposal due to the intractability of the transition density function. 
To alleviate this issue we have proposed to instead perform filtering for an approximate model where the approximation is such that it opens up for a more efficient particle filter proposal. This is in contrast with many existing approaches to approximate filtering, which are often based on approximating the inference algorithm itself. One motivation for this approach is that the model in most cases is an approximation to begin with, so adding a further approximation does not necessarily deteriorate the data analysis to any large degree. Another motivation is that an error analysis of the proposed procedure can focus on the model bias. Indeed, the inference method that we use is a regular particle filter (albeit for a non-standard model) and the properties of these methods are by know fairly well understood. 

% LORENZ ZO
\begin{figure*}[b]%
	\centering
	\begin{subfigure}{\columnwidth}
		\includegraphics[width=8.4cm]{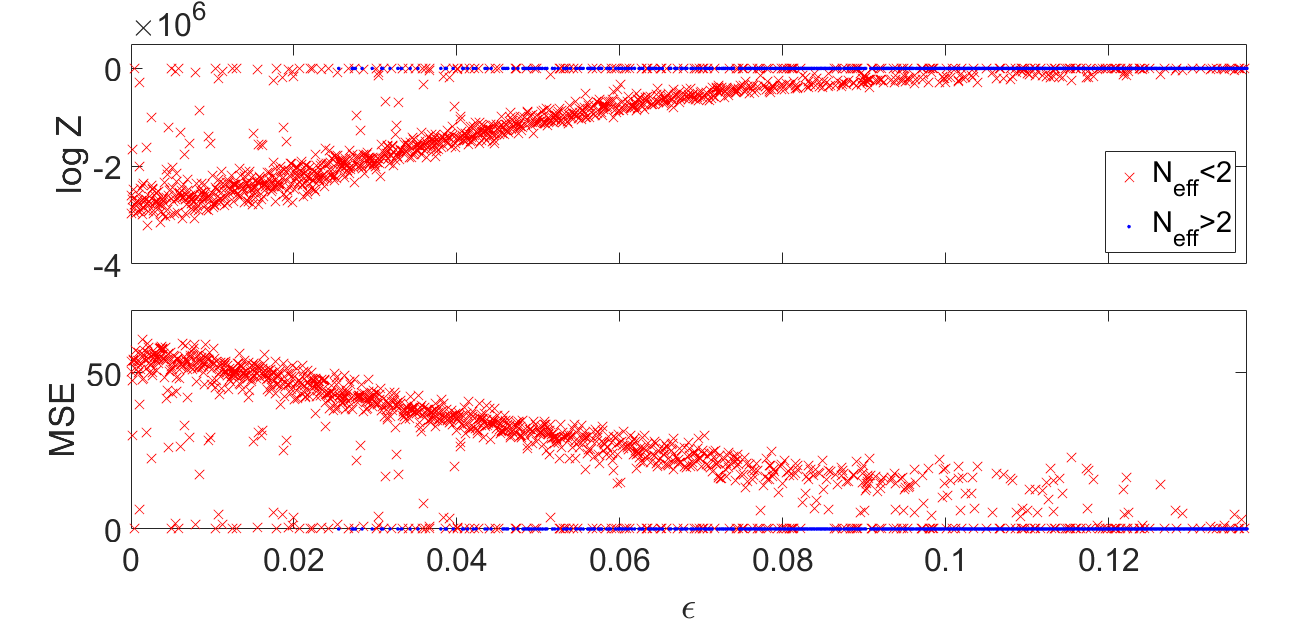}    
	\end{subfigure}\hfill%
	\begin{subfigure}{\columnwidth}
		\includegraphics[width=8.4cm]{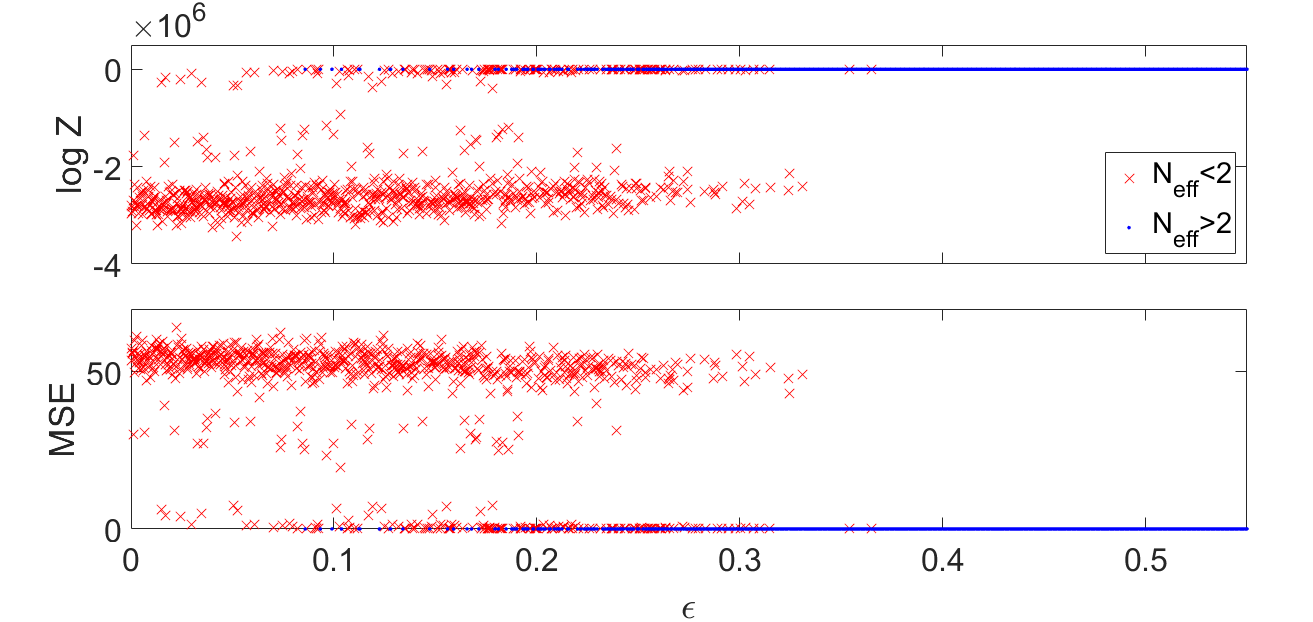}    
	\end{subfigure}\hfill%
	\caption{Marginal log-likelihood and MSE as a function of $\varepsilon$ for the 10-dimensional Lorenz'96 model. Left: Block-diagonal covariance matrix. Right: Weighted sample covariance matrix. Red crosses show estimates where the ESS drops below two at least once (indicating a degenerate filter estimate) and blue dots show estimates where the ESS is always greater than two.}
	\label{fig:LLHMSE_Lor_zo}
\end{figure*}

The proposed model approximation, and thus also the bias-variance trade-off of the resulting method, is controlled by the tuning parameter $\varepsilon$. From our numerical examples with two 10-dimensional state-space models it is clear that the introduction of a non-zero $\varepsilon$ can significantly improve the performance over the standard particle filter, both in terms of log-likelihood estimates and in terms of the filtering MSE. However, if we increase the dimension much further we expect that the proposed method will struggle as well. This is supported by the fact that even the locally optimal proposal will suffer from the curse of dimensionality \citep{Snyder15}. Thus, to address very high-dimensional problems (thousands of dimensions, say) we most likely need to combine the proposed method with other approaches. Such combinations of techniques is an interesting topic for future research.

So far we have only investigated the empirical performance of the new method for varying $\varepsilon$. In practice we would of course like to know which $\varepsilon$ to pick beforehand, or have an adaptive scheme for tuning $\varepsilon$ online. Devising such rules-of-thumb or adaptive methods is a topic for future work. The same hold for the choice of base covariance $S$ which needs further investigation. We note, however, that our empirical results suggest that the method is fairly robust to selecting $\varepsilon$ ``too large'', whereas a too small $\varepsilon$ resulted in very poor performance. A possible approach is therefore to start with a large $\varepsilon$ which is then gradually decreased while monitoring the ESS.

Finally, we note that the proposed method is useful not only for filtering problems, but also for system identification. Several state-of-the-art methods for identification of non-linear state-space models are based on the log-likelihood estimate (see, e.g., \citet{Schon15, Kantas15}). Thus, the significant improvement in these estimates offered by the proposed method should have direct bearing also on non-linear system identification.

\section*{Acknowledgements}
This research is financially supported by the Swedish Research Council via the project \emph{Learning of Large-Scale Probabilistic Dynamical Models} (contract number: 2016-04278) and by the Swedish Foundation for Strategic Research via the projects \emph{Probabilistic Modeling and Inference for Machine Learning} (contract number: ICA16-0015) and \emph{ASSEMBLE} (contract number: RIT15-0012).

\bibliographystyle{abbrvnat}
\bibliography{refs}

\appendix

\section{Additional plots}
Fig. \ref{fig:LLHMSE_Lor_zo} shows all runs for the 10-dimensional Lorenz'96 model. For small values on $\varepsilon$ the filter degenerates for both choices of covariance matrices. When $\varepsilon$ is increased the number of degenerate estimates gradually decreases.

\end{document}